\documentclass[english,12pt]{article}
\usepackage[T1]{fontenc}
\usepackage[latin1]{inputenc}
\usepackage{a4wide}
\usepackage{subfigure}
\usepackage{graphicx}
\usepackage{amssymb,cite}
\usepackage{psfrag}

\newcommand{\fract}[2]{{\textstyle\frac{#1}{#2}}}

\makeatletter

\providecommand{\LyX}{L\kern-.1667em\lower.25em\hbox{Y}\kern-.125emX\@}

\usepackage{babel}
\makeatother
\begin{document}
\makeatletter \renewcommand\theequation{\hbox{\normalsize\arabic{section}.\arabic{equation}}} \@addtoreset{equation}{section} \renewcommand\thefigure{\hbox{\normalsize\arabic{section}.\arabic{figure}}} \@addtoreset{figure}{section} \renewcommand\thetable{\hbox{\normalsize\arabic{section}.\arabic{table}}} \@addtoreset{table}{section} 

\makeatother

\title{{\normalsize \begin{flushright}\normalsize{ITP--Budapest Report 597}\end{flushright}\vspace{1cm}}
SUSY sine-Gordon theory as a perturbed conformal field theory and
finite size effects}

\author{Z. Bajnok$^{1}$, C. Dunning$^{2}$, L. Palla$^{3}$, G. Takács$^{1}$
and F. Wágner$^{3,4}$\\
{\normalsize $^{1}$Theoretical Physics Research Group, Hungarian
Academy of Sciences, Budapest, Hungary}\\
{\normalsize $^{2}$Department of Mathematics, University of York,
York, United Kingdom}\\
{\normalsize $^{3}$Institute for Theoretical Physics, Eötvös University,
Budapest, Hungary}\\
{\normalsize $^{4}$Department of Mathematics, King's College London,
  London, United Kingdom}}

\maketitle
\begin{abstract}
We consider SUSY sine-Gordon theory in the framework of perturbed
conformal field theory. Using an argument from Zamolodchikov, we
obtain the vacuum structure and the kink adjacency diagram of the
theory, which is cross-checked against the exact S matrix prediction,
first-order perturbed conformal field theory (PCFT), the NLIE method
and truncated conformal space approach. We provide evidence for
consistency between the usual Lagrangian description and PCFT on the
one hand, and between PCFT, NLIE and a massgap formula conjectured by
Baseilhac and Fateev, on the other. In addition, we extend the NLIE
description to all the vacua of the theory.
\end{abstract}

\section{Introduction}

Supersymmetric (SUSY) sine-Gordon theory (SSG) is of interest for
several reasons. First, it is a two-dimensional integrable field
theory and as such, there is a wealth of analytic and non-perturbative
information available about its behaviour. Second, it describes
supersymmetric solitons, which have been investigated recently in the
context of supersymmetric gauge theories in four (and other)
dimensions. It can be expected that SSG provides a useful laboratory
for the analysis of solitons in supersymmetric theories, and their
non-perturbative behaviour in general. Third, there has been a lot of
activity recently especially concerning the situation when SSG has a
nontrivial boundary condition imposed on a time-like boundary.

In spite of all the progress, much less is known about SSG than its
non-supersym\-metric counterpart, i.e. ordinary sine-Gordon
theory. Although its S matrix has been conjectured several years ago
\cite{ahn}, there is much less evidence for its correctness than in
the non-SUSY case. In particular, the issue of quantum corrections to
the soliton mass has been settled only fairly recently in \cite{BPS},
where it was proven that it satisfies a BPS-like property, even though
the theory only has $N=1$ supersymmetry.

There is also a certain controversy in the literature concerning the
vacuum and kink structure of the tricritical Ising model
\cite{fendley}, the simplest nontrivial SUSY integrable field theory,
which has a bearing on the S matrix of the SSG as well, since the S
matrix of tricritical Ising model provides the description of the
supersymmetric structure.

In this paper we set out to clarify the vacuum and kink structure of
the model (without boundaries), using information from several
sources. First, in Section 3 we apply an argument going back to
Zamolodchikov \cite{zam_susy} which gives a kink structure compatible
with the S matrix conjectured in \cite{ahn}. In Section 4 SSG is
considered in the perturbed conformal field theory framework, which is not
entirely trivial, as a purely bosonic potential term has to be omitted
from the Lagrangian (it is expected to be generated by radiative
corrections in this description). To show that this description is
correct we construct a spin-3 conserved charge and verify it against
the classical limit. In addition, we analyze first-order perturbative
corrections of the lowest-lying energy levels, and in Section 5
compare them to the results obtained from a conjectured NLIE, first
derived in \cite{CD}, which is extended to provide a finite volume
description of all the vacuum levels. Finally, in Section 6 we use the
truncated conformal space method to investigate the finite volume
spectrum.  Our conclusions are presented in Section 7.

\section{SSG theory}

\subsection{Action and discrete symmetries}

The supersymmetric sine-Gordon (SSG) theory is defined by the
action\begin{equation} 
\mathcal{A}_{\textrm{SSG}}=\int dtdx\, \left(\frac{1}{2}\partial _{\mu }\Phi \partial ^{\mu }\Phi +i\bar{\Psi }\gamma ^{\mu }\partial _{\mu }\Psi +m\bar{\Psi }\Psi \cos \frac{\beta }{2}\Phi +\frac{m^{2}}{\beta ^{2}}\cos \beta \Phi \right)\label{bulk_action}\end{equation}
where $\Phi $ is a real scalar, $\Psi $ is a Majorana fermion field,
$m$ is a mass parameter and $\beta $ is the coupling constant. The
theory is invariant under an $N=1$ supersymmetry algebra and has
infinitely many commuting local conserved charges \cite{ferrara}.
These charges survive at the quantum level and render the theory integrable,
which makes it possible to describe the exact spectrum and the S 
matrix. We use the Weyl representation for the spinor field\begin{eqnarray*}
\gamma ^{0} & = & \left(\begin{array}{cc}
 0 & i\\
 -i & 0\end{array}\right)\quad ,\quad \gamma ^{1}=\left(\begin{array}{cc}
 0 & i\\
 i & 0\end{array}\right)\quad ,\quad \gamma ^{3}=\gamma ^{0}\gamma ^{1}=\left(\begin{array}{cc}
 -1 & 0\\
 0 & 1\end{array}\right)\\
\Psi  & = & \left(\begin{array}{c}
 \psi _{-}\\
 \psi _{+}\end{array}\right)
\end{eqnarray*}
where $\psi _{\pm }$ are real Weyl components with definite chirality.

The supersymmetric theory has some discrete symmetries that play an
important role in what follows. The field theory interaction is periodic\begin{equation}
\Phi \rightarrow \Phi +n\frac{4\pi }{\beta },\quad n\in \mathbb{Z}\label{eq:periodic}\end{equation}
and even in the boson field\begin{equation}
\Phi \rightarrow -\Phi~. \label{eq:even}\end{equation}
These symmetries are also present in the non-supersymmetric version
of the sine-Gordon model. It is also interesting to observe that the Lagrangian is invariant
under a half-period shift \begin{equation}
\Phi \rightarrow \Phi +\frac{2\pi }{\beta }\label{eq:half-period}\end{equation}
if at the same time one changes the relative sign of the fermion components,
say\begin{equation}
\psi _{+}\rightarrow -\psi _{+}\quad ,\quad \psi _{-}\rightarrow \psi _{-}\label{eq:Kramers-Wannier}\end{equation}
or\[
\Psi \rightarrow -\gamma ^{3}\Psi~. \]

\subsection{Spectrum and scattering amplitudes}

The spectrum consists of the soliton/antisoliton multiplet, realizing
supersymmetry in a nonlocal way, and breathers that are bound states
of a soliton with an antisoliton. 

The building blocks of supersymmetric factorized scattering theory
were first described in \cite{schoutens}, using an Ansatz in which
the full scattering amplitude is a direct product of a part carrying
the SUSY structures and a part describing all the rest of the dynamics.
The full SSG S  matrix was constructed in \cite{ahn}.

The supersymmetric solitons are described by RSOS kinks $K_{ab}^{\epsilon }\left(\theta \right)$
of mass $M$ and rapidity $\theta $, where $a,\: b$ take the values
$0$, $\frac{1}{2}$ and $1$ with $|a-b|=1/2$, and describe the
supersymmetric structure, while $\epsilon =\pm $ corresponds to topological
charge $\pm 1$ (soliton/antisoliton). Multi-particle asymptotic states
are built as follows\begin{equation}
\left|K_{a_{0}a_{1}}^{\epsilon _{1}}\left(\theta _{1}\right)K_{a_{1}a_{2}}^{\epsilon _{2}}\left(\theta _{2}\right)\dots K_{a_{N-2}a_{N-1}}^{\epsilon _{N-1}}\left(\theta _{N-1}\right)K_{a_{N-1}a_{N}}^{\epsilon _{N}}\left(\theta _{N}\right)\right\rangle \label{rsos_sequence}\end{equation}
where $\theta _{1}>\theta _{2}>\dots >\theta _{N}$ for an \emph{in}
state and $\theta _{1}<\theta _{2}<\dots <\theta _{N}$ \emph{}for
an \emph{out} state. The two-particle scattering process \[
K_{ab}^{\epsilon _{1}}\left(\theta _{1}\right)+K_{bc}^{\epsilon _{2}}\left(\theta _{2}\right)\, \rightarrow \, K_{ad}^{\epsilon '_{2}}\left(\theta _{2}\right)+K_{dc}^{\epsilon '_{1}}\left(\theta _{1}\right)\]
has an amplitude of the form\begin{equation}
S_{\textrm{SUSY}}\left(\left.\begin{array}{cc}
 a & d\\
 b & c\end{array}\right|\theta _{1}-\theta _{2}\right)\times S_{SG}\left(\theta _{1}-\theta _{2}\, ,\, \lambda \right)_{\epsilon _{1}\epsilon _{2}}^{\epsilon '_{1}\epsilon '_{2}}\label{bulk_ampl}\end{equation}
i.e. the tensor structure of the scattering amplitude factorizes into
a part describing the SUSY structure (which we call the SUSY factor)
and another part corresponding to the topological charge (the bosonic
factor).

The bosonic factor coincides with the usual sine-Gordon S  matrix,
but the relation between the parameter $\lambda $ and $\beta $ is
different from the sine-Gordon case\[
\lambda =\frac{8\pi }{\beta ^{2}}-\frac{1}{2}~.\]
The SUSY factor is identical to the S  matrix of the tricritical
Ising model perturbed by the primary field of dimension $\frac{6}{5}$
\cite{zam_susy}:\begin{eqnarray}
\renewcommand{\arraystretch}{1.25}
 &  & S_{\textrm{SUSY}}\left(\left.\begin{array}{cc}
 0 & \frac{1}{2}\\
 \frac{1}{2} & 0\end{array}\right|\theta \right)=S_{\textrm{SUSY}}\left(\left.\begin{array}{cc}
 1 & \frac{1}{2}\\
 \frac{1}{2} & 1\end{array}\right|\theta \right)=2^{(i\pi -\theta )/2\pi i}\cos \left(\frac{\theta }{4i}-\frac{\pi }{4}\right)K(\theta )\nonumber \\
 &  & S_{\textrm{SUSY}}\left(\left.\begin{array}{cc}
 \frac{1}{2} & 0\\
 0 & \frac{1}{2}\end{array}\right|\theta \right)=S_{\textrm{SUSY}}\left(\left.\begin{array}{cc}
 \frac{1}{2} & 1\\
 1 & \frac{1}{2}\end{array}\right|\theta \right)=2^{\theta /2\pi i}\cos \left(\frac{\theta }{4i}\right)K(\theta )\nonumber \\
 &  & S_{\textrm{SUSY}}\left(\left.\begin{array}{cc}
 0 & \frac{1}{2}\\
 \frac{1}{2} & 1\end{array}\right|\theta \right)=S_{\textrm{SUSY}}\left(\left.\begin{array}{cc}
 1 & \frac{1}{2}\\
 \frac{1}{2} & 0\end{array}\right|\theta \right)=2^{(i\pi -\theta )/2\pi i}\cos \left(\frac{\theta }{4i}+\frac{\pi }{4}\right)K(\theta )\nonumber \\
 &  & S_{\textrm{SUSY}}\left(\left.\begin{array}{cc}
 \frac{1}{2} & 1\\
 0 & \frac{1}{2}\end{array}\right|\theta \right)=S_{\textrm{SUSY}}\left(\left.\begin{array}{cc}
 \frac{1}{2} & 0\\
 1 & \frac{1}{2}\end{array}\right|\theta \right)=2^{\theta /2\pi i}\cos \left(\frac{\theta }{4i}-\frac{\pi }{2}\right)K(\theta )\nonumber \\
 &  & K(\theta )=\frac{1}{\sqrt{\pi }}\prod _{k=1}^{\infty }\frac{\Gamma \left(k-\frac{1}{2}+\frac{\theta }{2\pi i}\right)\Gamma \left(k-\frac{\theta }{2\pi i}\right)}{\Gamma \left(k+\frac{1}{2}-\frac{\theta }{2\pi i}\right)\Gamma \left(k+\frac{\theta }{2\pi i}\right)}~.\label{eq:tricritical_scattering}
\end{eqnarray}
The supersymmetry factor has no poles in the physical strip and so
the full supersymmetric amplitudes of the kinks have the same poles
as in the ordinary sine-Gordon model. The corresponding bound states
are supersymmetric breathers $B_{n}$ of mass\begin{equation}
m_{n}=2M\sin \frac{\pi n}{2\lambda }\quad ,\quad n=1,\dots ,[\lambda ]\label{eq:mass_spectrum}\end{equation}
which transform in the ordinary boson-fermion doublet representation
of the SUSY algebra. The bosonic component of $B_{n}$ has a nontrivial
$(-1)^{n}$ parity under bosonic field reflection $\Phi \rightarrow -\Phi $.

\section{Kink and vacuum structure of SUSY models}

Here we use an argument by Zamolodchikov \cite{zam_susy} to clarify
the vacuum and kink structure of SUSY sine-Gordon theory. In its
original form the argument is for the tricritical Ising model; we
briefly recall it here.

\subsection{Tricritical Ising model \protect \\
}

In the bosonic Landau-Ginzburg formulation, the basic field is the
(bosonic) spin field $\sigma $, and the action is of the form\[
\int \left(\frac{1}{2}\left(\partial _{t}\sigma \right)^{2}-\frac{1}{2}\left(\partial _{x}\sigma \right)^{2}-V(\sigma )\right)dxdt~,\]
where the potential has the form\[
V\left(\sigma \right)=\lambda \left(\sigma ^{2}-\sigma _{0}^{2}\right)^{2}\sigma ^{2}~.\]

\psfrag{s}{$\sigma$}
\begin{figure}
\begin{center}\includegraphics{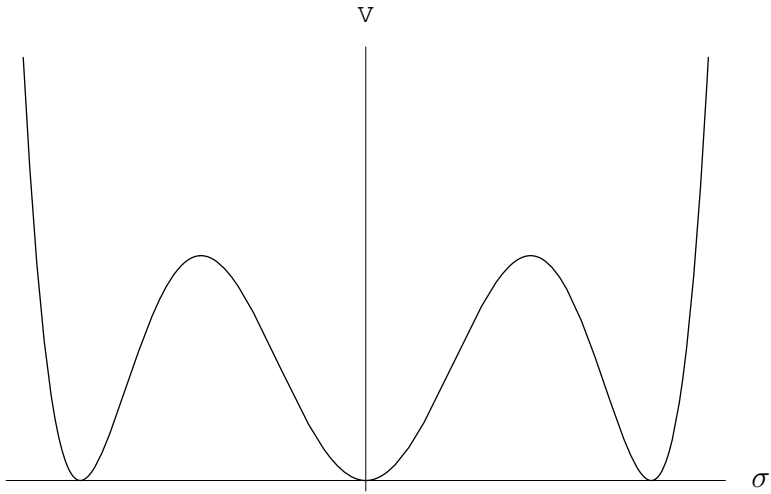}\end{center}

\caption{\label{potential-bosonic} The scalar potential in the Landau-Ginzburg
description of the tricritical Ising model}
\end{figure}

\begin{figure}
\begin{center}\includegraphics{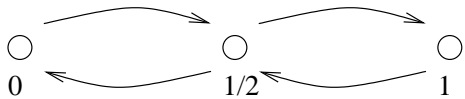}\end{center}

\caption{\label{adj-tricritical} The adjacency diagram of kinks in the
  tricritical Ising model}
\end{figure}

\noindent As  can be seen from the form of the potential (Fig. \ref{potential-bosonic}),
there are three vacua in the model, and the kinks connect them according
to the adjacency diagram in Fig. \ref{adj-tricritical}, with a scattering
amplitude given in (\ref{eq:tricritical_scattering}).

\begin{figure}
\begin{center}\includegraphics{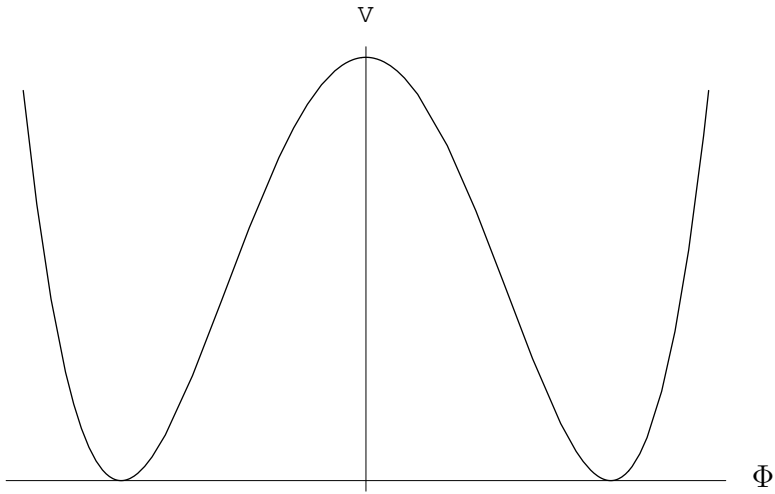}\end{center}

\caption{\label{potential-susy} The scalar potential in the SUSY LG description
of the tricritical Ising model}
\end{figure}

However, in the supersymmetric Landau-Ginzburg formalism the action 
(in component form) looks like 
\[
\int \left(\frac{1}{2}\left(\partial _{t}\Phi
  \right)^{2}-\frac{1}{2}\left(\partial _{x}\Phi \right)^{2}+\psi
  _{+}\partial _{-}\psi _{+}+\psi _{-}\partial _{+}\psi _{-}+W'(\Phi
  )\psi _{+}\psi _{-}+W(\Phi )^{2}\right)dxdt~, 
\]
where 
\[ W(\Phi )=\mu \left(\Phi ^{2}-\lambda ^{2}\right)~.
\]
The scalar potential is of the form (Fig\.{ } \ref{potential-susy})\[
V(\Phi )=\mu ^{2}\left(\Phi ^{2}-\lambda ^{2}\right)^{2}\]

\psfrag{f}{$\Phi$}

\noindent and has two minima \[
\Phi =\pm \lambda~. \]
How do we reconcile this picture with the bosonic one, in which there
are three vacua? Zamolodchikov's argument goes as follows: at the
minimum $\Phi =\lambda $, the fermion mass is positive, there is
only one vacuum, the Majorana fermion describes the high temperature
phase of the Ising model and the vacuum expectation value of the spin
field vanishes:\[
\langle \sigma \rangle=0~.\]
However, when $\Phi =-\lambda $, we are in the low temperature phase
of the Ising model. Therefore this vacuum is twofold degenerate, with\[
\langle \sigma \rangle =\pm \sigma _{0}~.\]
 The kinks go from the $\langle \sigma \rangle=0$ vacuum to the
 $\langle\sigma \rangle=\pm \sigma _{0}$ 
ones. Therefore there are two pairs of kinks, and we get the same
adjacency diagram (Fig.~\ref{adj-tricritical}), with the
$\langle\sigma \rangle=0$ 
(RSOS label $\frac{1}{2}$) vacuum in the middle, while the other two 
$\langle\sigma \rangle=\pm \sigma _{0}$ (RSOS labels $0,1$) sit at the edges.

It must be mentioned that Fendley has conjectured another bootstrap
S  matrix for the tricritical Ising model in \cite{fendley}, based
on an assumption that the kink structure in the SUSY Landau-Ginzburg
formulation was different from that of the bosonic formulation. The
bosonic and the super Landau-Ginzburg formulations are two different
local descriptions of the model; their relation is very similar to
the sine-Gordon/massive Thirring connection \cite{heretic}. However,
in the case of the sine-Gordon/massive Thirring relation the kink
S  matrix is independent of the local description taken, and Zamolodchikov's
argument indicates that this is the case for the tricritical Ising
model as well. As a result, in both the bosonic and the fermionic
description the kink scattering is described by the same RSOS S 
matrix (\ref{eq:tricritical_scattering}), contrary to Fendley's assumption.%
\footnote{TBA results also show that Fendley's S  matrix yields an irrational
value for the ultraviolet central charge, thus it cannot correspond
to a perturbation of tricritical Ising model (M. Moriconi, private
communication).%
} The difference between the local algebras manifests itself in the
different boundary conditions prescribed for the local fields, just
as in the sine-Gordon/massive Thirring case (cf. also \cite{FRTodd}
for the situation in finite volume).

\subsection{N=1 SUSY sine-Gordon theory}

\begin{figure}
\begin{center}\input{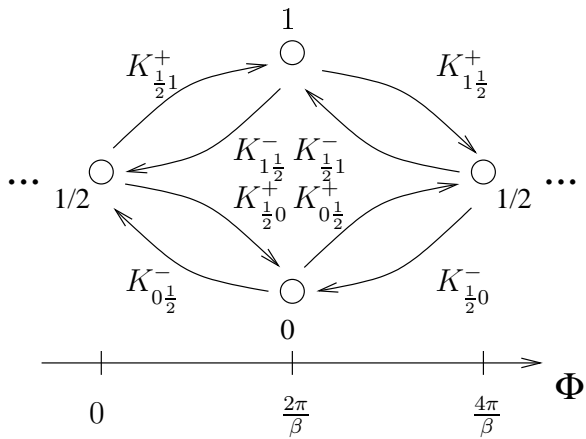}\end{center}
\caption{\label{adjacency-SG} A period of the adjacency diagram of kinks
in supersymmetric sine-Gordon theory}
\end{figure}

Now let us apply these considerations to $N=1$ SUSY sine-Gordon theory.
Here the bosonic potential is\[
-\cos \beta \Phi \]
while the mass term of the fermion is of the form\[
\psi _{-}\psi _{+}\cos \frac{\beta \Phi }{2}~.\]
As a result, every even vacuum $\Phi =2n\frac{2\pi }{\beta }\, ,\, n\in Z$
is nondegenerate, while every odd vacuum $\Phi =(2n+1)\frac{2\pi }{\beta }\, ,\, n\in Z$
is doubly degenerate, with nonzero expectation value of the spin field
associated to the Majorana fermion. We have a vacuum structure analogous
to that of tricritical Ising model, which periodically repeats itself
(similarly to the periodic vacuum structure of the sine-Gordon model).
This results in the adjacency diagram shown in Fig.~\ref{adjacency-SG},
which is exactly the one that describes the tensor product S  matrix
of kinks $K_{ab}^{\pm }\; ,\; a=0,\frac{1}{2},1$.

The discrete symmetries of the Lagrangian are manifest on the adjacency
diagram, which is periodic in $\Phi $ with period $4\pi /\beta $.
The transformation (\ref{eq:half-period}, \ref{eq:Kramers-Wannier})
is a translation with $2\pi /\beta $, together with the analogue
of Kramers-Wannier duality acting on the Majorana fermion, and exchanges
the broken phase of the fermion (two vacua) and the symmetric phase
(a single vacuum).

Due to the periodicity of the vacuum structure and the field theoretic
potential, in finite volume one can create twisted sectors depending
on a $\vartheta $ angle according to \cite{polymer}, or one can 
define folded versions of the model by identifying vacua after a certain
number of periods \cite{folded}.

\subsection{The IR description}

Now we give a qualitative description of the consequences of imposing
periodic boundary conditions in the SSG theory in the IR domain, i.e
in a large but finite volume $L$. It is based on the known features
of the spectrum and the exact S  matrix. 

The first consequence of imposing periodic boundary conditions is
that the single kink states are excluded. Some of the two kink states
are allowed; namely those $K_{ab}K_{bc}$ where the $a$ and $c$
vacua are identical. As two kinks can fuse into breathers one may
have single breather states in the sectors where the topological charge
vanishes. It is easy to see, using the $a\rightarrow 1-a$ symmetry
of the SUSY factor (\ref{eq:tricritical_scattering}), that the allowed
two kink states are divided into two sets, 
\[
(K_{0\frac{1}{2}}K_{\frac{1}{2}0},\, \, K_{1\frac{1}{2}}K_{\frac{1}{2}1})\qquad \qquad (K_{\frac{1}{2}0}K_{0\frac{1}{2}},\, \, K_{\frac{1}{2}1}K_{1\frac{1}{2}})\]
closed under scattering. The first set contains two vacua (the ones
indexed by $0$ and $1$), while the second one has a single vacuum
(the one indexed by $\frac{1}{2}$). It is a natural expectation that
in a finite volume these three vacua are no longer degenerate but
their energies are exponentially close to each other. Above all three
vacua we also expect the repetition of the spectrum of (single) breather
states albeit with their energies split. One of the aims of this paper
is to give a qualitative verification of this picture.

The two sets of periodic two kink states have rather different supersymmetry
properties. To establish this we use the kink-kink-breather fusion
coefficients \[
|K_{ab}(\theta +i\xi /2)K_{bc}(\theta -i\xi /2)\rangle =f_{abc}^{\phi }|\phi (\theta )\rangle +f_{abc}^{\psi }|\psi (\theta )\rangle ,\]
 (where $|\phi \rangle $ and $|\psi \rangle $ stand for the bosonic
and fermionic breathers respectively) explicitly given in \cite{HolM}.
These explicit expressions reveal that the two kink states in the
first set can fuse into bosonic breathers only, while the ones in
the second set can fuse into both bosonic and fermionic breathers.
Furthermore, using the action of supersymmetry charges (also given
in \cite{HolM}), one can show that there is an unbroken $N=1$ supersymmetry
in the space of states of the second set.

\section{SSG as a perturbed conformal field theory}

\subsection{The perturbed conformal field theory picture \label{sub:The-perturbed-conformal}}

We consider SUSY sine-Gordon theory as a perturbation of the $c=3/2$
conformal field theory of a free boson and a free fermion. The Hilbert
space of the theory is taken to be a tensor product of the conformal
free boson and free fermion, and we consider only local sectors, i.e.
retain only states which have integer conformal spin. At generic values
of the compactification radius this permits only two boundary conditions
on the fermion field\[
\Psi (x+L)=\pm \Psi (x)\]
which means either Neveu-Schwarz or Ramond boundary conditions on
both components of the fermion fields (different boundary conditions
on the components would result in sectors of noninteger spin).

We take a conformal boson field $\phi $ compactified on a circle
of radius $r$, normalized with the action\[
\frac{1}{8\pi }\int d^{2}z\partial \phi \bar{\partial }\phi \]
which means that the canonical boson field of (\ref{bulk_action})
can be written as $\Phi =\frac{1}{\sqrt{4\pi }}\phi $. The conformal
boson has a $\hat{U}(1)\times \hat{U}(1)$ symmetry generated by the
modes of the currents \[
J(z)=i\partial \phi=\sum _{n}a_{n}z^{-n-1}\quad \textrm{and}\quad
\bar{J}(\bar{z})=i\bar{\partial }\phi=\sum
_{n}\bar{a}_{n}\bar{z}^{-n-1}~.\] 
 The highest weight vectors $|n,m\rangle $ of the current algebra
are created by the vertex operators \[
V_{(n,m)}(z,\bar{z})=:e^{i(\frac{n}{r}+\frac{mr}{2})\varphi (z)+i(\frac{n}{r}-\frac{mr}{2})\bar{\varphi }(\bar{z})}:~.\]
In order to have a maximal local operator algebra we take both $n$
and $m$ to be integers. The Hamiltonian, on the cylinder of circumference
$L$, in the sector $|n,m\rangle $ has the following form
\[
H_{boson}=\frac{2\pi }{L}\left(\left(\frac{n}{r}\right)^{2}+\left(\frac{mr}{2}\right)^{2}+\sum _{k>0}(a_{-k}a_{k}+\bar{a}_{-k}\bar{a}_{k})-\frac{1}{12}\right)~,\]
 where the bosonic modes satisfy\[
[a_{k},a_{l}]=k\delta _{k+l}\quad ;\quad [a_{k},\bar{a}_{l}]=0\quad ;\quad [\bar{a}_{k},\bar{a}_{l}]=k\delta _{k+l}~.\]

In the Neveu-Schwarz sector, the fermionic part of the Hilbert space
can be generated by the negative modes of the Euclidean fermionic
fields (corresponding to the Minkowski fields $\psi _{\pm }$) \[
\psi (z)=\sum _{s+1/2\in Z}b_{s}z^{-s-1/2}\quad ;\quad \bar{\psi }(\bar{z})=\sum _{s+1/2\in Z}\bar{b}_{s}\bar{z}^{-s-1/2}\]
acting on the Neveu-Schwarz vacuum $|NS\rangle $ and locality requires
the total fermion number to be even. The modes satisfy

\[
\{b_{s},b_{t}\}=\delta _{s+t}\quad ;\quad \{b_{s},\bar{b}_{t}\}=0\quad
;\quad \{\bar{b}_{s},\bar{b}_{t}\}=\delta _{s+t}~.\] 
The fermionic Hamiltonian takes the form (using conformal normalization
conventions)\[
H_{NS}=\frac{1}{2\pi }\int d^{2}z\, \psi \bar{\partial }\psi
=\frac{2\pi }{L}\left(\sum _{s=\frac{1}{2}}^{\infty
  }b_{-s}b_{s}-\frac{1}{24}\right)~.\] 
In the Ramond sector, the fermionic field has integer mode expansion
and both the left and the right fermion number must be even. We denote
the highest weight state by $|R\rangle $. The Hamiltonian in the
Ramond sector is \[
H_{R}=\frac{2\pi }{L}\left(\sum _{s=1}^{\infty
  }b_{-s}b_{s}+\frac{1}{48}\right)~.\] 
Writing $\cos \frac{\beta }{2}\Phi =\frac{1}{2}(V_{(1,0)}+V_{(-1,0)})$
gives the relation $\frac{1}{r}=\frac{\beta }{4\sqrt{\pi }}$
between  the compactification radius $r$ and the coupling $\beta $,
and this leads to \[
p\equiv \frac{1}{\lambda }=\frac{2}{r^{2}-1}~.\]

\subsection{Conserved charges in the PCFT framework and the perturbing potential}

A convenient way to describe the SUSY sine-Gordon  model is to
consider it as an appropriate perturbation of the conformal field
theory in its UV limit \cite{Zamp} by the operator $U_{1}=\bar{\psi }\psi \cos \frac{\beta \Phi }{2}$.
In \cite{Mansf} the SSG is described as a perturbation of the super
Liouville theory and it is shown that the purely bosonic piece of
the potential $U_{2}=\frac{m^{2}}{\beta ^{2}}\cos \beta \Phi $ vanishes
in the renormalized theory. 

The problem is that assuming the kinetic terms of the boson and fermion
fields in the Lagrangian provide the description of the $c=\frac{3}{2}$
theory, the two parts of the interaction $U_{1}=\bar{\psi }\psi \cos \frac{\beta \Phi }{2}$
and $U_{2}$ have different conformal dimensions: $\Delta _{1}=\bar{\Delta }_{1}=\frac{1}{2}+\frac{\beta ^{2}}{32\pi }$
and $\Delta _{2}=\bar{\Delta }_{2}=\frac{\beta ^{2}}{8\pi }$. This,
and the fact that they have different supersymmetry properties would
complicate the PCFT description. Nevertheless -- using the conserved
quantities -- we argue below that for the SSG it is enough to consider
only $U_{1}$ as a perturbation. 

This perturbation obviously preserves a $(1,1)$ supersymmetry generated by
the supercurrents $G$ and $\bar{G}$, since in the superfield formalism
$U_{1}$ can be written as $G_{-1/2}\bar{G}_{-1/2}\cos (\frac{\beta }{2}\hat{\Phi })$
with $\cos (\frac{\beta }{2}\hat{\Phi })$ being a Neveu-Schwarz superconformal
primary field \cite{Ger}. Thus the real question is whether $U_{1}$
provides an integrable perturbation of the $c=\frac{3}{2}$ theory.
For this it is enough if $U_{1}$ preserves a single higher spin conserved
quantity. 

One can look for a conserved spin $3$ quantity generated by the density
$T_{4}$, where \[
T_{4}=(\partial _{z}^{2}\Phi )^{2}+A(\partial _{z}\Phi )^{4}+B(\partial _{z}\Phi )^{2}\partial _{z}\psi \psi +C\partial _{z}^{2}\psi \partial _{z}\psi \, ,\]
 with constants $A$, $B$ and $C$. In the first order of the perturbation
$T_{4}$ is conserved if in the operator product $T_{4}(z)U_{1}(w,\bar{w})$
the residue of the first order pole is a total derivative \cite{Zamp}.
Having computed the residue one can eliminate the terms containing
the derivatives of the $\psi $ field by using total derivatives;
the remaining (non total derivative) terms then have the form \[
(\tilde{A}\psi \partial _{z}^{3}\Phi +\tilde{B}\psi \partial _{z}^{2}\Phi \partial _{z}\Phi +\tilde{C}\psi (\partial _{z}\Phi )^{3})\cos (\frac{\beta }{2}\Phi )\, .\]
 Demanding the vanishing of $\tilde{A}$, $\tilde{B}$ and  $\tilde{C}$
yields a linear inhomogeneous system of equations for $A$, $B$,
$C$, that has a solution, which shows that $U_1$ generates an
integrable perturbation. Furthermore the $\beta \rightarrow 0$
limits of $A$, $B$, $C$ reproduce the classical expressions obtained
in \cite{Inam} using both $U_{1}$ and $U_{2}$ as perturbations
together with the classical equations of motion. This provides the
justification for only using  $U_{1}$ as the perturbation in the PCFT
framework. 

Therefore in terms of the canonically normalized fields the Lagrangian
of the SSG is written in the perturbed CFT framework as \[
\mathcal{L}=\frac{1}{2}\partial _{\mu }\Phi \partial ^{\mu }\Phi +i\bar{\Psi }\gamma ^{\mu }\partial _{\mu }\Psi +\mu \bar{\Psi }\Psi \cos \frac{\beta }{2}\Phi \, .\]
 The coefficient of the perturbing potential is denoted here by $\mu $
to emphasize that its ($\beta $ dependent) dimension is different
from that of the classical mass $m$. Identifying the bulk SUSY sine-Gordon
model with the $n=2$ case of \cite{BF} gives the following relation
between the kink mass $M$ and the $\mu $ parameter \begin{equation}
\frac{\mu }{8}\gamma \left(\frac{1}{2}-\frac{\beta ^{2}}{32\pi }\right)=M^{1-\frac{\beta ^{2}}{16\pi }}\left(\frac{\pi }{4}\frac{\beta ^{2}}{16\pi -\beta ^{2}}\right)^{1-\frac{\beta ^{2}}{16\pi }},\quad \gamma (x)=\frac{\Gamma (x)}{\Gamma (1-x)}\, .\label{eq:massgap}\end{equation}
 This relation - which connects an IR and a UV parameters and is called
the massgap relation - plays an important role in writing the TCSA program
devised to analyze the finite volume spectrum of SSG. Note that for
$\beta \rightarrow 0$, the massgap relation correctly reproduces
the classical kink (soliton) mass $M=\frac{8\mu }{\beta ^{2}}$.

\subsection{The PCFT analysis of energy levels in the UV}

The first PCFT correction to the UV energy levels can be expressed in
terms of certain integrals over the complex plane of various 2p and 4p
functions in the underlying conformal field theory \cite{KM}. Indeed
writing the (bare) Euclidean action of the perturbed CFT on the
cylinder as 
\[ {\mathcal{A}}_{g
}={\mathcal{A}}_{\textrm{CFT}}+g \int d^{2}\xi \, \chi (\xi ),\]
where $g$ has mass dimension $y=2-2\Delta _{\chi }$ and $\chi
(z)$ is normalized by $\langle \chi (z,\bar{z})\chi (0,0)\rangle
=|z|^{-4\Delta _{\chi }}$, for small dimensionless volumes $l=LM$, the
first correction to the energy of a state $|a\rangle $, with
$L_{0}|a\rangle =\Delta _{a}|a\rangle =\bar{L}_{0}|a\rangle $, can be
written as \[ \frac{6l}{\pi }E_{a}(l)=-(c-24\Delta
_{a})-c_{2}^{a}\kappa ^{2}l^{2y}+{\texttt {o}}(l^{4y}).\] Here $c$ is
the central charge of the UV CFT, it is assumed that the symmetry
properties of the fields forbid any first order (${\texttt
{o}}(l^{y})$) corrections and to derive this formula the massgap
relation is written in the form $g =\kappa M^{y}$. We remark
that the bulk energy constant in SSG vanishes on account of the
supersymmetry. The coefficient of the leading correction at
$l\rightarrow 0$ can be written explicitly as \begin{equation}
c_{2}^{a}=6(2\pi )^{1-2y}\int \limits
_{\textrm{plane}}\frac{d^{2}z}{|z|^{y}}\langle a|\chi (1,1)\chi
(z,\bar{z})|a\rangle |_{\textrm{conn}}\, .\label{c2}\end{equation}

Using the conformal normalized free fermions $\psi $, $\bar{\psi }$
and the properly normalized $V_{(\pm 1,0)}$ vertex operators to describe
the perturbation in the SUSY sine-Gordon model one obtains \[
\chi (z,\bar{z})=\bar{\psi} (\bar{z})\psi (z)\frac{1}{\sqrt{2}}(V_{(1,0)}(z,\bar{z})+V_{(-1,0)}(z,\bar{z}))\]
and 
\begin{equation}
g =\frac{\mu }{2\pi \sqrt{2}}~.\label{mu-lambda}
\end{equation}
 This perturbation is relevant if the compactification radius $r>1$,
and then $|0\rangle $, $|1\pm \rangle $, $|2\pm \rangle $ and $|3\pm \rangle $
constitute most of the lowest lying states in the various NS sectors.
(Here $|N\pm \rangle $, $N=1,2,3$ denote the states 
$|N\pm \rangle =\frac{1}{\sqrt{2}}(V_{(N,0)}(0,0)|{0}\rangle \pm V_{(-N,0)}(0,0)|{0}\rangle )$
with $|0\rangle $ being the tensor product of the bosonic
and NS vacua $|{0}\rangle =|\tilde{0}\rangle \otimes |NS\rangle $).
For these states it is straightforward to determine the conformal
correlation functions needed in eq.(\ref{c2}), and the integrals
can be evaluated using the general expression \begin{equation}
\int d^{2}z\frac{|z|^{\rho -2}}{|1-z|^{2\nu }}=\pi \gamma \left(\nu -\frac{\rho }{2}\right)\gamma \left(\frac{\rho }{2}\right)\gamma (1-\nu )\, .\label{eq:regularization}\end{equation}
 (The integral is convergent at $z\rightarrow 0$ if $\rho >0$, at
$z\rightarrow 1$ if $\nu <1$ and at $z\rightarrow \infty $ if
$2\nu -\rho >0$; we keep the conditions $2\nu >\rho >0$ but continue
in $\nu $ above $\nu =1$ using the r.h.s. if necessary). The coefficients
of the leading PCFT corrections for these seven states are summarized
in Table \ref{cap:pct_table}, where the coefficient $\alpha $ is
\[
\alpha =\frac{3}{2}(8)^{\frac{2}{r^{2}}}\frac{1}{\gamma ^{2}\left(\frac{1}{2}-\frac{1}{2r^{2}}\right)}\left(\frac{1}{r^{2}-1}\right)^{2-\frac{2}{r^{2}}}\, .\]

\begin{table}
\renewcommand{\arraystretch}{1.25}
\begin{center}\begin{tabular}{|c|c|c|}
\hline 
State&
 $\Delta _{a}$&
 $c_{2}^{a}\kappa ^{2}$\\
\hline
$|0\rangle $&
 0&
 $\alpha \gamma ^{2}\left(\frac{1}{2}+\frac{1}{2r^{2}}\right)\gamma \left(-\frac{1}{r^{2}}\right)$\\
\hline
$|1\pm \rangle $&
 $\frac{1}{2r^{2}}$&
 $\alpha \gamma (\frac{1}{2}-\frac{1}{2r^{2}})\left[\gamma (-\frac{1}{r^{2}})\gamma (\frac{1}{2}+\frac{3}{2r^{2}})\pm \frac{1}{2}\gamma (\frac{1}{2}-\frac{1}{2r^{2}})\gamma (\frac{1}{r^{2}})\right]$\\
\hline
$|2\pm \rangle $&
 $\frac{2}{r^{2}}$&
 $\alpha \gamma (\frac{1}{2}-\frac{3}{2r^{2}})\gamma (-\frac{1}{r^{2}})\gamma (\frac{1}{2}+\frac{5}{2r^{2}})$\\
\hline
$|3\pm \rangle $&
 $\frac{9}{2r^{2}}$&
 $\alpha \gamma (\frac{1}{2}-\frac{5}{2r^{2}})\gamma (-\frac{1}{r^{2}})\gamma (\frac{1}{2}+\frac{7}{2r^{2}})$ \\
\hline
\end{tabular}\end{center}

\caption{\label{cap:pct_table}The first PCFT corrections for some states}
\end{table}

Note that in all these cases it is necessary to make the analytical
continuation in $\nu $ to give meaning to the otherwise divergent
integrals. This divergence is expected as the conformal dimension
of the perturbing operator $\Delta =\frac{1}{2}+\frac{1}{2r^{2}}$ is
greater than $1/2$; 
thus the singularity coming at $z\rightarrow 1$ from the $\chi (1)\chi (z)$
OPE is not integrable no matter what the states $|a\rangle $
are. 

In the Ramond sector we expect that the ground state scaling function
vanishes identically as a result of unbroken supersymmetry. The question
is whether the (leading) PCFT corrections are consistent with this
expectation. The conformal contribution to the scaling function vanishes
since\[
c-24\Delta _{R}=\frac{3}{2}-24\frac{1}{16}=0~.\]
The next contribution is proportional to the second order coefficient
\[
c_{2}^{|R\rangle }=6(2\pi )^{1-2y}\int \limits _{\textrm{plane}}\frac{d^{2}z}{|z|^{y}}\langle R|\chi (1,1)\chi (z,\bar{z})|R\rangle |_{\textrm{conn}}\, ,\]
where\[
|R\rangle =\lim _{z,\bar{z}\rightarrow 0}\sigma (z,\bar{z})|NS\rangle \]
denotes the Ramond ground state. The integrand defining $c_{2}^{|R\rangle }$
is formally positive, but the integral is divergent and needs regularization.
Separating the integrand into a sum of terms, after an appropriate
partial integration it can be converted into a form in which each
term can be regularized using (\ref{eq:regularization}). The sum
of regularized terms turns out to vanish. In this sense PCFT is consistent
with a vanishing ground state scaling function in the Ramond sector.

\section{NLIE}
To probe the behaviour of the
scaling functions between the ultraviolet and infrared limits 
one can use the thermodynamic Bethe Ansatz \cite{yang,ZamTBA}.
Fendley and Intriligator's TBA gluing idea \cite{FI} leads to 
a set of TBA equations for all choices of the SSG coupling
\cite{RTV1,RTson,RT}, but in practice the equations take a simple form only at 
certain $\beta$. Instead one can 
treat all values of the coupling on an equal 
footing by using an alternative type of nonlinear
integral equation, originally developed for the ground state of the
sine-Gordon model in \cite{DDV} (and independently in a related
context in \cite{KP,KBP}),  
and which usually goes by the acronym NLIE.

The equations proposed in \cite{CD} for the SSG model consist of a
TBA-like function 
coupled to a nonlinear integral equation. The two pieces reflect the 
factorized nature of the S matrix, with the TBA part associated to the 
SUSY factor and the NLIE part to the bosonic factor. 
The original equations of \cite{CD} generate the lowest
ground state in the 
 Neveu-Schwarz sector, but a simple sign change and
 appropriate choice of parameters provides access to the other two
 ground states. 
The equations are 
\begin{eqnarray*}
\ln y_1(\theta) &=& -il \sinh (\theta)+i\pi\omega   +\int_{-\infty}^{\infty} 
\!d\theta'\,\chi(\theta{-}\theta'{+}\fract{i\pi}{2})\ln (1 + (-1)^{\delta}
y_2(\theta'))  \nonumber \\
&&\quad +\int_{{\cal C}_1} 
\!d\theta'\,\varphi(\theta{-}\theta')\ln (1+y_1(\theta')) -
\int_{{\cal C}_2} 
\!d\theta'\,\varphi(\theta{-}\theta')\ln (1+y_1^{-1}(\theta')) ~;
\\  [11pt]
\ln y_2(\theta) &=& \int_{{\cal C}_1} 
\!d\theta'\,\chi(\theta{-}\theta'{-}\fract{i\pi}{2})\ln (1+y_1(\theta')) -
\int_{{\cal C}_2} 
\!d\theta'\,\chi(\theta{-}\theta'{-}\fract{i\pi}{2})\ln (1+y_1^{-1}(\theta'))~.
\end{eqnarray*}
The integrations contours ${\cal C}_1$ and ${\cal C}_2$ run from
$-\infty $ to $\infty$ just below and above the real axis
respectively. 
The 
kernel $\varphi(\theta)$ is   proportional to the logarithmic
derivative of the
soliton-soliton scattering amplitude of the sine-Gordon model
\[
\varphi(\theta)=\int \frac{dk}{2\pi}\,e^{i k\theta}
\frac{\sinh (p{-}1)\frac{\pi k}{2}}
{2\sinh \frac{\pi\,p\, k}{2} \cosh \frac{\pi k}{2} }~,
\label{sgker}
\]
and $\chi(\theta)$ is related to the SUSY factor of
the S matrix 
\[
\chi(\theta)=1/(2\pi\cosh \theta) .
\]
The exact ground state energy depends only on the NLIE-like function via
$y_1(\theta)$ 
\[
E(l)= -\frac{i}{2 \pi }
\left( \int_{{\cal C}_1}d\theta\, \sinh \theta \, \ln
  (1+y_1(\theta)) 
-\int_{{\cal C}_2}d\theta\, \sinh \theta \,\ln
(1+y_1^{-1}(\theta))\right)~.
\]

By setting the parameters $\omega$ and $\delta$ appropriately these
equations provide access to all three vacua, though  not for all
possible choices of $(\beta,l)$ for each state, as we shall comment on
further below. The  
appropriate settings are shown in table~\ref{cap:nlie}.
\begin{table}
\renewcommand{\arraystretch}{1.25}
\begin{center}\begin{tabular}{|l|c|c|}
\hline 
State & $\omega$ & $\delta$ \\
\hline
$|\tilde{0}\rangle \otimes |NS\rangle $&  $0$ & $0$ \\  \hline
$|1+\rangle$& $1$ & $0$ \\  \hline
$|\tilde{0}\rangle \otimes |R\rangle $& $\frac{1}{2}$ & $1$ \\
\hline
\end{tabular}
\end{center}
\caption{\label{cap:nlie}
Choice of NLIE parameters to obtain each of the three ground states.}
\end{table}

Typically we can extract  a
 number of results 
 analytically from  nonlinear integral equations, the simplest of
 which is the ultraviolet value of the ground state energy. In the 
Neveu-Schwarz sector ($\delta=0$) the scaling function behaves as : 
\[
\frac{6l}{\pi} E (0) = -\left (\frac{3}{2} - \frac{12\omega^2}{r^2}
\right )~,
\]
while in the Ramond sector $(\delta=1)$ we have
\[
\frac{6l}{\pi} E (0) = -\left (- \frac{12  (\omega-1/2)^2}{r^2}
\right )~.
\]
Note that 
tuning $\omega$ according to Table~\ref{cap:nlie} yields the correct
UV behaviour for each ground state:
\[
\frac{6l}{\pi} E (0) = -\left ( c-24 \Delta_a \right)~.
\]
Furthermore the conformal dimension of the
perturbing operator exactly matches that of $U_1$.  
In the infrared limit ($l \to
\infty$) we deduce the  scaling functions behaviour to be \cite{CD} 
\begin{equation}
E(l) \sim -2 \sqrt 2 \cos \pi \omega \, \int \frac{d\theta}{2\pi} \cosh\theta \, e^{-Ml\cosh\theta}~.
\label{IRbehavior}\end{equation}
This result can be given a simple intuitive interpretation. First, let
us note that the RSOS structure describing the SUSY degrees of freedom
of solitons has three nodes $0$, $1/2$, $1$, which correspond to
truncating the quantum group representation theory ${\cal U}_q(sl(2))$
at $q^4=1$ to the tensor product rule
\begin{eqnarray*}
&&0\otimes a=a\quad ,\quad a=0,\frac{1}{2},1\\
&&\frac{1}{2}\otimes\frac{1}{2}=0\oplus 1\quad ,\quad 1\otimes 1=0~.
\end{eqnarray*}
There is a notion of statistical dimension for these representations
$d_a$ ($a=0,1/2,1$), which satisfies the ordinary rules under the
tensor product
\begin{eqnarray*}
&&d_0d_a=d_a\qquad ,\qquad a=0,\frac{1}{2},1\\
&&d_{\frac{1}{2}}^2=d_0 + d_1\qquad ,\qquad d_1^2=d_0~,
\end{eqnarray*}
and the solution to these conditions is 
\[
d_0=d_1=1 \qquad , \qquad d_{\frac{1}{2}}=\sqrt{2}~.
\]
The kinks correspond to the $1/2$ representation, and the above
results mean that they have a statistical dimension $\sqrt{2}$. It
might seem strange since it is hard to interpret this as the multiplet
length. However, a simple calculation shows that the number of
$2n$-kink states (neglecting for the moment the topological charge)
with periodic boundary conditions grows as $2^n$ (there are no
odd-kink states on a circle) so the kinks indeed behave as a multiplet
of length $\sqrt{2}$. The reason for this can be found in the
nontrivial vacuum adjacency conditions imposed on multi-kink states,
which do not allow for all possible sequences to be realized.

Armed with this, it is easy to interpret the result
(\ref{IRbehavior}). For $\omega=0$ it is exactly the leading
correction from $2\sqrt{2}$ particles in finite volume (or
equivalently finite temperature) to the free energy. The additional
factor $2$ comes from the fact that the topological charge introduces
a further doublet structure of the kinks. The $\omega$ dependence is
easy to understand using the same sort of instanton argument as in the
case of the ordinary sine-Gordon model \cite{folded}.
\footnote{We remark that a similar explanation can be made for the IR
behaviour of the vacuum scaling function predicted by the NLIE, for
all the fractional supersymmetric sine-Gordon models, c.f. formula
(29) in \cite{CD}. For the case $L$ of that paper, the statistical
dimension of the kink representation $1/2$ results in
$2\cos(\pi/(L+2))$. SSG corresponds to $L=2$, and ordinary sine-Gordon
to $L=1$ (i.e. the kinks have dimension $1$ apart from the double
degeneracy corresponding to their topological charge).}

A final check can be made by comparing  NLIE data
with the above perturbed conformal  theory predictions.  
Turning to the first Neveu-Schwarz ground state, we extracted 
the coefficient $b_2$ 
from NLIE data fitted  to a suitably truncated series of the form  
 \[
\frac{6l}{\pi }E_{a}(l)=-(c-24\Delta _{a})- \sum_{n=1} b_{2n} l^{2ny}~,
\]
then made a comparison with the coefficient of the leading PCFT correction
$c_2^a\kappa^2$. The results displayed in
Table \ref{cap:nlie_table} show excellent agreement.  
\begin{table}
\renewcommand{\arraystretch}{1.25}
\begin{center}\begin{tabular}{|c| c |c|}
\hline 
$r$&
 $b_2$&
 $c_{2}^{a}\kappa ^{2}$\\
\hline
$\sqrt{15}/3 $ & $-0.4305230886667 $ &  $-0.4305230886637$\\ \hline
$\sqrt{21}/3$ & $ -0.7099942272385$ & $  -0.7099942272382$ \\ \hline
  $2$ & $   -0.5922281367467$ & $  -0.5922281367458$ \\ \hline 
$3$ & $    -0.2517882823698$ & $  -0.2517882823567$ \\ \hline
$3.05$ &$-0.2422117898616$ & $-0.2422117898994$ \\ \hline
\end{tabular}\end{center}
\caption{\label{cap:nlie_table}Comparison of PCFT correction with NLIE for
  the first Neveu-Schwarz ground state}
\end{table}     
Numerical iteration of the NLIEs in the Ramond sector confirms  
the Ramond ground state energy is identically zero for
all choices of $l$.  This provides a nontrivial check of the equations since  
the NLIE functions $y_1(\theta)$ and $y_2(\theta)$ are not
identically zero.  

\begin{figure}
\begin{center}\includegraphics[%
width=9cm]
{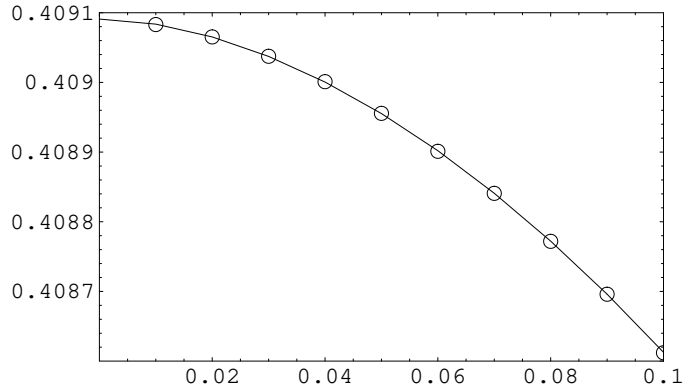}
\end{center}
\caption{\label{fig:nlie} Comparing PCFT with NLIE for the second
  Neveu-Schwarz ground state}
\end{figure}

It turns out to be harder to check the NLIEs for the second
Neveu-Schwarz ground state. For small values of $r$ and large values of
the cylinder size $l$ the equations work well.  This allows a
comparison with PCFT, but does not allow one to obtain the value of $c_2^a k^2$
to as many decimal places as for the lower NS ground state. 
Instead, in Fig.~\ref{fig:nlie} we compare the PCFT 
result at compactification radius $r^2=11$ (equivalently $p=1/5$)   
\[  c-24\Delta_{|1+\rangle}+c_2^{|1+\rangle} k^2 l^{2y}~, \]
using a solid line 
to the NLIE data $-6lE_a(l)/\pi$ (indicated by  
 the symbol  $ \circ $). The agreement is very good. 

So what lies behind the difficulties at some couplings and/or cylinder
size?
For larger values of $r$ or smaller $l$ the equations fail because
one or more
zeros of $1+y_1(\theta)$ attempt to cross one of the integration
contours, resulting in a singularity in $\ln (1+y_1(\theta))$. (Once
this has happened a zero of $1+y_2(\theta)$ may also attempt to cross
a contour.) 
This behaviour is not unexpected: by sending the cylinder size $l$ into the
complex plane  the authors of \cite{DTan1,DTan2} were able to 
analytically continue a set of TBA
equations describing the ground state energy of a particular model  to
a new set, which  
 gave access to the first excited state. Here, we have an additional
 parameter $\omega$ which we   
analytically continue  from $0$ to $1$ along the real axis to obtain
the second NS ground state from the first.   
 Singularities in the logarithmic term of NLIEs 
of the type discussed here also arise via a second mechanism, 
which usually occurs as  the
 (real) cylinder size $l$ is decreased and is not  explicitly
connected to an excited state.
 This so-called `specials' problem 
was first described in the context of the  ordinary sine-Gordon NLIE
when the equation was tuned to  
 study the various states of the minimal models perturbed by $\phi_{13}$
\cite{FRT3}.  Since we are able to make a favourable comparison of
NLIE data and PCFT results for some couplings $r$ 
we leave the resolution of these problems for 
future investigation.  This will also provide clues as how to modify the
NLIEs to address the excited states,  as has been done for the
sine-Gordon NLIE in \cite{FMQR,DDV2,FRT2,FRT1,FRTodd}.

\section{TCSA}

\subsection{TCSA for the supersymmetric sine-Gordon model} 

The supersymmetric sine-Gordon model can be viewed as a relevant
perturbation of the $c=\frac{3}{2}$ free conformal field theory,
consisting of a free fermion and a free boson. Following the idea of
the truncated conformal space approach (TCSA) \cite{TCSA}, the Hilbert
space is truncated at a given conformal energy $E_{\textrm{cut}}$, and so the
Hamiltonian can be diagonalized numerically to obtain an approximate
finite volume spectrum. Since the perturbing operator has scaling
dimension larger than $1/2$ we are faced with divergences in the TCSA
data and hope to obtain relative energy levels only. Furthermore, the
Hilbert space contains many more vectors up to a given energy cut than
in a theory with a single scalar field only, so only qualitative
results are expected.

The matrix elements of the perturbing operator between any two conformal
states $|a\rangle $ and $|b\rangle $ with conformal weights $(\Delta _{a},\bar{\Delta }_{a})$
and $(\Delta _{b},\bar{\Delta }_{b})$ can be brought to the following
form
\begin{eqnarray*}
\langle a|g \!\!\int _{0}^{L}\!dx\, \chi (\xi ,\bar{\xi
})|b\rangle  \!\!\!& =\!\!\! & \frac{g L}{\sqrt{2}}\left(\frac{2\pi
  }{L}\right)^{2\Delta } \!\!\langle a|i\psi (1)\bar{\psi
}(1)\left(V_{(1,0)}(1,1)+V_{(-1,0)}(1,1)\right)|b\rangle \, \delta
_{\Delta _{a}-\bar{\Delta }_{a},\Delta _{b}-\bar{\Delta }_{b}} 
\end{eqnarray*}
 using a conformal mapping from the cylinder to the plane and translational
invariance (the $\delta $ expresses the conservation of momentum/conformal
spin). Recall that $\Delta =\frac{1}{2}+\frac{1}{2r^{2}}$.

We use relations (\ref{eq:massgap},\ref{mu-lambda}) to write the PCFT coupling
$g$ in terms of the kink mass $M$ and to convert the Hamiltonian
into dimensionless form (energy measured in units of $M$), parametrized
by the dimensionless volume parameter $l=ML$. The Hilbert space can
be decomposed into sectors of given conformal spin and given bosonic
winding number (topological charge).

Using the mode expansions and $U(1)$ charge conservation, the matrix
elements of the perturbing operator can be calculated as the product
of separate bosonic and fermionic contributions. In the Ramond sector,
matrix elements including the fermion zero mode can be computed using
the relation
\[
\langle R|ib_{0}\bar{b}_{0}|R\rangle =\frac{1}{2}~.\]
Since the interaction is even in the bosonic field, the perturbing
operator has the following discrete bosonic $Z_{2}$ symmetry
\[
|n,m\rangle\leftrightarrow |-n,-m\rangle \quad ;\quad a_{n}\leftrightarrow
-a_{n}\quad ;\quad \bar{a}_{n}\leftrightarrow -\bar{a}_{n}~.\] 
 The discrete symmetry (\ref{eq:half-period},\ref{eq:Kramers-Wannier})
means that the NS sector has an additional $Z_{2}$ symmetry acting
as 
\[
|n,m\rangle \leftrightarrow (-1)^{n}|n,m\rangle \quad ;\quad
b_{s}\leftrightarrow -b_{s}\quad ,\, \bar{b}_{s}\leftrightarrow
\bar{b}_{s}~.\] 
There is no such symmetry in the Ramond sector as a result of the presence
of the fermionic zero mode. 

The Hilbert space (see Section \ref{sub:The-perturbed-conformal}) can
be further decomposed with respect to these symmetries. In table \ref{cap:firststates}
we list the lowest energy states of each sector for zero spin and
zero topological charge. The first sign refers to the NS parity, while
the second to the bosonic one. In parenthesis we indicate the symbol
we used in the TCSA data for the various sectors and 
$|N\pm\rangle_\textrm{R}=\frac{1}{\sqrt{2}}\left(V_{(N,0)}(0,0)|0\rangle_\textrm{R}
\pm V_{(-N,0)}(0,0)|0\rangle_\textrm{R}\right)$, 
where $|0\rangle_\textrm{R}=|\tilde{0}\rangle\otimes|R\rangle$. 

\begin{table}
\renewcommand{\arraystretch}{1.15}
\begin{center}\begin{tabular}{|c|c|c|c|c|c|}
\hline 
$++\quad \, {\scriptstyle (\sqcup \! \! \! \! \sqcap )}$&
$-+\quad \, {\scriptstyle (\circ )}$&
$+-\quad \, {\scriptstyle (\times )}$&
$--\quad \, {\scriptstyle (+\hspace {-6.5pt}\times )}$&
$+\quad \, {\scriptstyle (\bigtriangleup )}$&
$-\quad \, {\scriptstyle (+)}$\\
\hline
\hline 
$|0\rangle$&
$|1+\rangle$&
&
&
$|0\rangle_\textrm{R}$&
\\
\hline 
&
&
$|2-\rangle$&
$|1-\rangle$&
&
$|1-\rangle_\textrm{R}$\\
\hline 
$|2+\rangle$&
$|3+\rangle$&
&
&
$|1+\rangle_\textrm{R}$&
\\
\hline
\end{tabular}\end{center}

\caption{\label{cap:firststates} UV classification of the lowest lying
  states 
in the various NS and R sectors for $r>3$. For $r<3$ the order of
the states is somewhat different ($|3+\rangle$
is the third state in its sector), but they can still be identified
unambiguously.}
\end{table}

\subsection{Checking TCSA against PCFT}

The aim of this investigation is to confirm the TCSA program by comparing
the small volume (UV) data for the low lying energy levels to the
predictions of conformal perturbation theory (PCFT). Since the TCSA data
are given in terms of the (IR) kink mass while the PCFT predictions
in terms of the (UV) parameter $\mu $, to make the formal comparison
one has to use the massgap relation. Nevertheless as the TCSA program
is written using the same massgap relation the success of the comparison
says nothing about the correctness of this relation but may confirm
the TCSA program. 

Since the scaling dimension of the perturbing operator is larger than
$1/2$ both the conformal perturbation theory and TCSA are plagued
by divergences. As discussed in section 3.3, in PCFT these divergences
are regularized by analytical continuation. The introduction of a
finite $E_{\textrm{cut}}$ in TCSA also regularizes these divergences
albeit in a different way. Therefore the best one can do to compare
TCSA and the PCFT predictions is to consider the differences between
the various energy levels rather than the levels themselves: \[
\frac{6l}{\pi }(E_{b}(l)-E_{a}(l))-24(\Delta _{b}-\Delta _{a})=\Delta c^{ba}\kappa ^{2}l^{2y}+{\texttt {o}}(l^{4y}),\quad \Delta c^{ba}=c_{2}^{a}-c_{2}^{b}\, .\]
 This has two advantages: on the one hand the TCSA energy differences
depend less sensitively on $E_{\textrm{cut}}$ than the individual
levels, and on the other the integrals defining $\Delta c^{ba}$ are
convergent thus they may be computed as the appropriate differences
of the data in the table \ref{cap:pct_table}. 

\begin{figure}
\begin{center}\subfigure[r=1.9]{\includegraphics[scale=0.9]{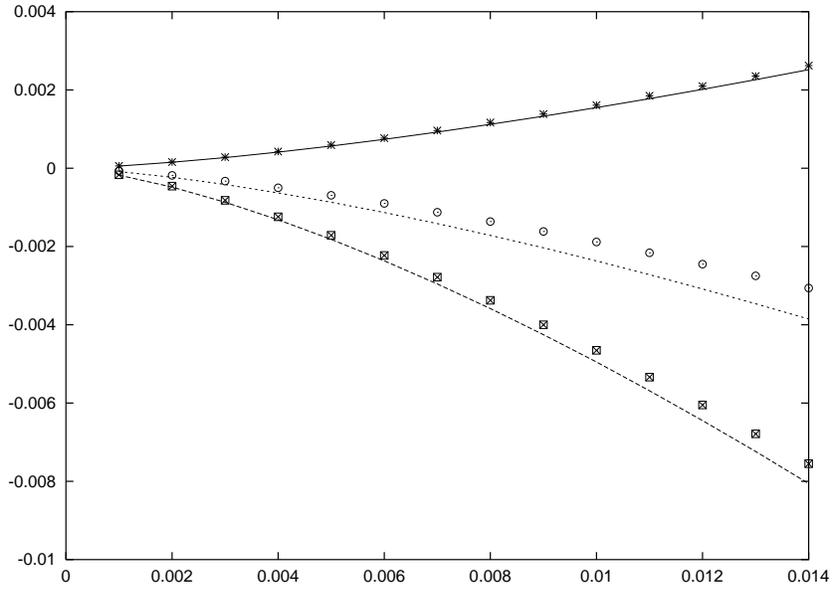}}\end{center}

\begin{center}\subfigure[r=2.3]{\includegraphics[scale=0.9]{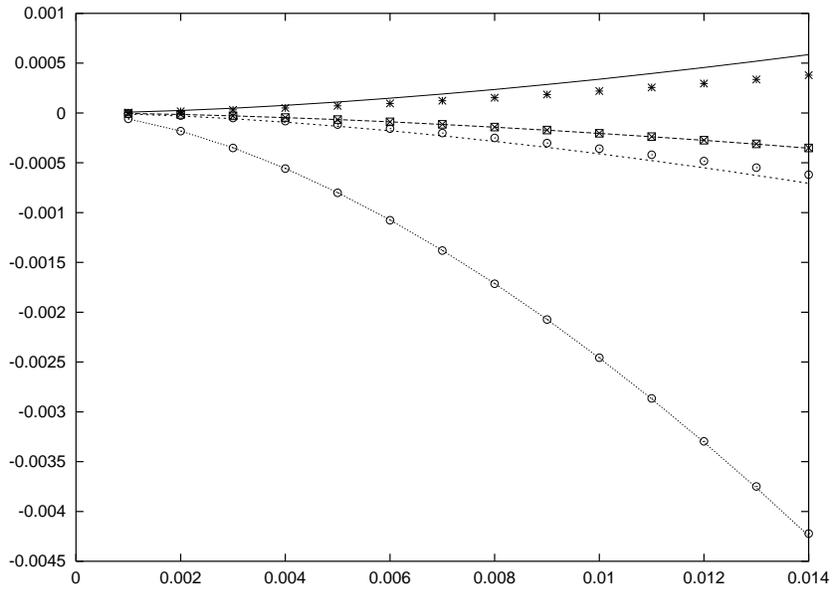}}\end{center}
\caption{\label{cap:Comparison-of-TCSA} Comparison of TCSA and PCFT energies
relative to the NS (++) vacuum}
\end{figure}

Using the identifications between the low lying TCSA lines and the
$|N\pm \rangle $ states in table \ref{cap:firststates} we compare
the TCSA energy differences and the PCFT predictions in Fig.~\ref{cap:Comparison-of-TCSA}.
In diagram (a), the three continuous lines depict the PCFT predictions
with $\Delta c^{|1-\rangle |0\rangle }$ $\Delta c^{|1+\rangle |0\rangle }$
and $\Delta c^{|2+\rangle |0\rangle }$ respectively, while - in accordance
with table \ref{cap:firststates} - the symbols $*$, $\circ$, $\square $
and $\times $ denote the TCSA data for \begin{eqnarray*}
 &  & \left(E_{1}^{--}(l)-E_{1}^{++}(l)\right)\frac{6l}{\pi }-\frac{12}{r^{2}}\quad ,\\
 &  & \left(E_{1}^{-+}(l)-E_{1}^{++}(l)\right)\frac{6l}{\pi }-\frac{12}{r^{2}}\quad ,\\
 &  & \left(E_{2}^{++}(l)-E_{1}^{++}(l)\right)\frac{6l}{\pi }-\frac{48}{r^{2}}\quad \mathrm{and}\\
 &  & \left(E_{1}^{+-}(l)-E_{1}^{++}(l)\right)\frac{6l}{\pi }-\frac{48}{r^{2}}\quad ,
\end{eqnarray*}
respectively. Note that the data is consistent with the $|2\pm \rangle $
states being degenerate in the leading order of PCFT.

In diagram (b), the first three lines and the first four sets of TCSA
data are just like on (a), while the fourth continuous line depicts
the leading PCFT prediction for $\Delta c^{|3+\rangle |1+\rangle }$
and the second line of points marked by $\circ$ are the TCSA data \[
\left(E_{3}^{-+}(l)-E_{1}^{-+}(l)\right)\frac{6l}{\pi }-\frac{96}{r^{2}}~.\]
The agreement between the data and the leading predictions is excellent,
though the deviations between them indicate that in some cases the
higher order PCFT corrections are not negligible even in this $l$
region.

\subsection{IR spectrum from TCSA}

\begin{figure}
\begin{center}\includegraphics{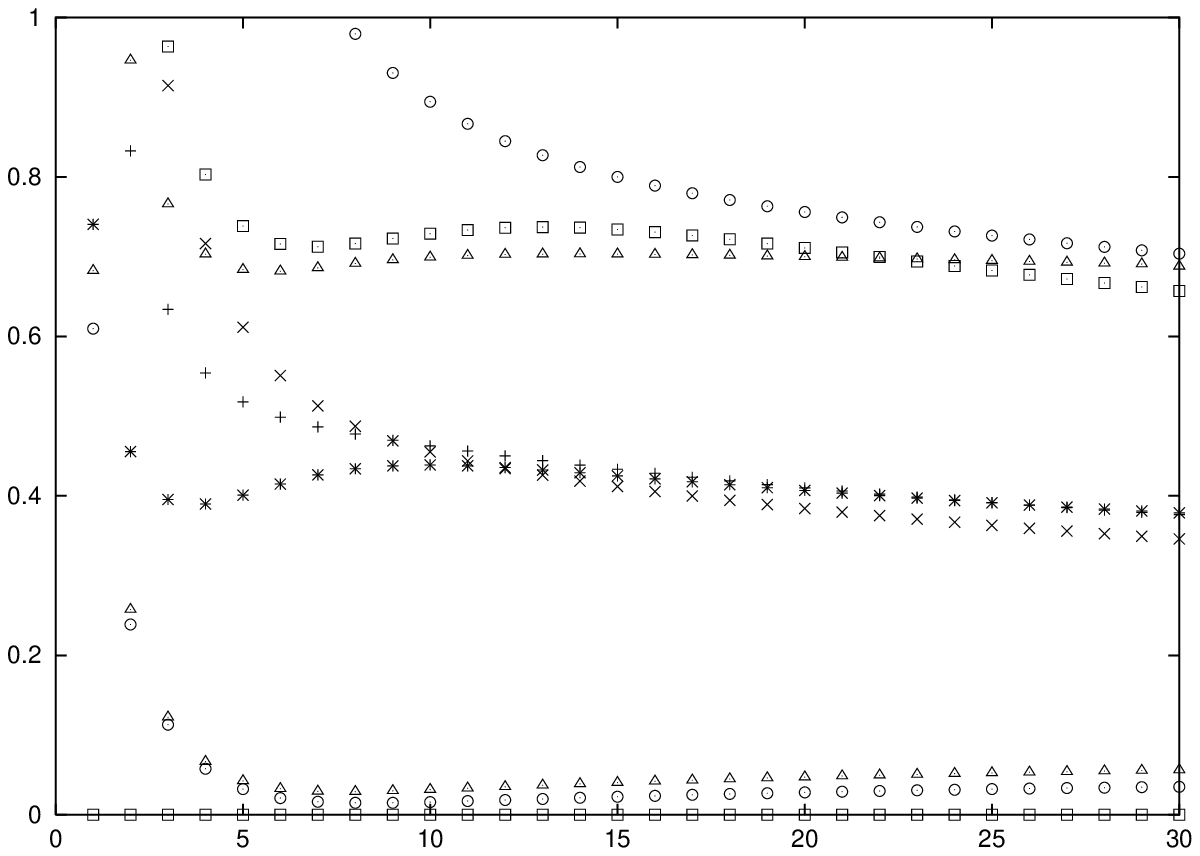}\end{center}

\caption{\label{cap:tcsa_spec} Lowest lying states in the TCSA spectrum}
\end{figure}

Since the scaling dimension of the perturbing operator is larger than
$1/2$ we have divergences in the TCSA data. This can be seen by increasing
the cutoff: the energy levels become more and more negative. The energy
differences, however, converge but unfortunately very slowly. The
first two lines in the $(++)$, $(-+)$ NS and in the $(+)$ R sectors,
and the first (lowest) lines in the $(+-)$, $(--)$ NS and $(-)$
R sectors are shown in Fig.~\ref{cap:tcsa_spec}, where we plot these
lines relative to the ground state of the $(++)$ NS sector. 

The various points are marked according to table \ref{cap:firststates}.
The first row in the table corresponds to the UV limit of the three
vacua as shown on the data. The next row, (the next three curves)
corresponds to the first breathers excited above each vacua, while
the last row, (the last three curves) are related to the second breathers.
The bosonic breathers are identified on the basis of their behaviour
under the discrete symmetries. The compactification radius is $r=3.05$
and the energy cut is $14$, which corresponds to approximately $13000$ states
in the Ramond sector and $9000$ states in the NS sectors, respectively.
Clearly the curves are in a qualitative agreement with the spectrum,
but the particular mass values are wrong. Indeed it can be seen that
the TCSA converges very slowly to the right values: by changing the
truncation level from $8$ to $14$ in steps of $2$, the energy differences
change almost the same amount in each step. However, the qualitative
features of the spectrum (i.e. the degeneracy and gap patterns) already
appear for very small number of states and are stable against increasing
the truncation level. 

Even though the mass values converge very slowly, the case for the mass
ratios is much better. Calculating particular mass ratios from the
data, the predictions of S  matrix theory (\ref{eq:mass_spectrum})
can be reproduced with a few percent accuracy, and are much more stable
against changing the cut.

Furthermore, the TCSA data are also qualitatively consistent with
the dependence of the spectrum on the coupling, i.e. by changing the
compactification radius $r$ the number and relative position of
breather line triplets follows the predicted spectrum.

\section{Conclusions}

In this work we investigated supersymmetric sine-Gordon theory (SSG),
especially the structure of the vacua and the kinks connecting
them. Using an argument due to Zamolodchikov we clarified the vacuum
and kink structure and showed that it corresponds exactly to the exact
S matrix conjectured in \cite{ahn}. Then we formulated the theory in
the perturbed conformal field theory (PCFT) framework, which shows 
a
striking difference to the usual Lagrangian description, namely the
omission of a purely scalar potential term. The first evidence we gave
for the correctness of this description came from the construction of
a spin-3 conserved charge, which is necessary for
integrability. Considering the classical limit, we showed that the
expression for this charge obtained in the PCFT framework reduces to
the result known from the classical Lagrangian approach, evidence
that the two approaches indeed describe the same model.

Using PCFT, we obtained the leading behaviour of some energy levels in
finite volume, for the limit of small volume ( the ultraviolet regime). In
particular, we concluded that the energy of the Ramond ground state
vanishes in this limit, which is consistent with the fact that it is
expected to vanish exactly due to unbroken SUSY in finite volume in
this sector. 

The PCFT results were then compared to results obtained from an NLIE
equation for the lowest lying Neveu-Schwarz state, which is the true
ground state in finite volume. This comparison involves the mass gap
relation between the PCFT coupling (an UV parameter) and the soliton
mass (an IR quantity) conjectured earlier in \cite{BF}. The comparison
showed  excellent agreement for several different values of the
dimensionless coupling parameter $\beta$ (equivalently $r$) of the
model, which is a strong evidence for the correctness of the mass gap
formula on one hand, and for the consistency between the PCFT and NLIE
approach on the other. 

We have also examined the large volume (infrared) limit of the NLIE
and found a perfect agreement with considerations based on statistics
of kinks and instanton calculus in finite volume. 

The NLIE proposed in \cite{CD} was formulated for the true ground
state of SSG, which is the lowest lying state in the Neveu-Schwarz
sector. Matching conformal dimensions computed from the NLIE with
those predicted from conformal field theory, we proposed equations
describing the other two vacua, one in the Ramond and the other in the
Neveu-Schwarz sector. Numerical iteration of the NLIE for the Ramond
ground state gave a vanishing energy for all volumes, as 
expected, providing a very strong check since the solution of the
equation itself is still nontrivial in this case. For the second
Neveu-Schwarz vacuum, convergence properties of the iteration are not
as good as for the other two cases, but the UV behaviour matches the
PCFT prediction nicely, giving another confirmation of the consistency
between PCFT, NLIE and the massgap formula.

Finally, we performed an analysis of the finite volume problem using
truncated 
conformal space approach (TCSA). We checked the validity of the
approach in the UV regime by a comparison with PCFT. While the
numerical convergence of the TCSA is not very good (partially due to
UV divergences, but also to a rapidly growing number of states as the
energy cut is raised), it does confirm the qualitative picture of the
spectrum, both the degeneracies and the behaviour of the breather
levels are in accordance with the picture presented before.

There are many open issues. It would be very interesting to give a
complete characterization of the SSG finite volume spectrum from the
NLIE, like it has been done in the non-SUSY case
\cite{KP,DDV,FMQR,DDV2,FRT1,FRT2,FRTodd,FRT3}. It is not at all
obvious how to achieve that, however, mainly because the NLIE for the
SSG theory is of a mixed type - part of the system resembles the
Destri-de Vega equation, while the other part looks like a thermodynamic
Bethe Ansatz equation. It is likely that a reformulation of the NLIE is
necessary.

Another interesting problem would be to extend the analysis of the
spectrum to the boundary case, along the lines of non-SUSY sine-Gordon
theory \cite{BSG}. This was part of the motivation to introduce TCSA
for SSG, however, the slow convergence of TCSA prevents us from
obtaining reliable quantitative results, which would be necessary to
check the bootstrap spectrum proposed in \cite{BSSG}. It remains to be
seen whether this can be overcome in some way. Extending the NLIE to
the boundary situation could be another way to obtain interesting
results and check the bootstrap predictions.

\subsection*{Acknowledgments}

The authors would like to thank Gerard Watts for illuminating
discussions. We are also grateful to Marco Moriconi for sharing with
us his unpublished TBA results on Fendley's scattering theory. TCD
thanks the UK EPSRC for a Research Fellowship. This work was partially
supported by the EC network ``EUCLID'', contract number
HPRN-CT-2002-00325, and Hungarian research funds FKFP 0043/2001, OTKA
D42209, T037674, T034299 and T043582. GT was also supported by a
Széchenyi István Fellowship, and ZB by a Bolyai János Research
Fellowship.


\begin{thebibliography}{10}

\bibitem{ahn}C. Ahn: 
\emph{Nucl. Phys.} \textbf{B354} (1991) 57-84.

\bibitem{BPS}
H.~Nastase, M.~A.~Stephanov, P.~van Nieuwenhuizen and A.~Rebhan:
\emph{Nucl. Phys.} \textbf{B542} (1999) 471-514, hep-th/9802074.\\
N.~Graham and R.~L.~Jaffe,
\emph{Nucl. Phys.} {\bf B544} (1999) 432-447, hep-th/9808140.\\
M.~A.~Shifman, A.~I.~Vainshtein and M.~B.~Voloshin,
\emph{Phys. Rev.} {\bf D59} (1999) 045016, hep-th/9810068.\\
A.~Rebhan, P.~van Nieuwenhuizen and R.~Wimmer,
\emph{Nucl. Phys.} {\bf B648} (2003) 174-188, hep-th/0207051.

\bibitem{fendley}P. Fendley: \emph{Phys. Lett.} \textbf{B250} (1990) 96-101.

\bibitem{zam_susy}A.B. Zamolodchikov: 
\texttt{Fractional-spin integrals of motion in perturbed conformal
field theory}, in Fields, Strings and Quantum Gravity, eds. H. Guo, Z. Qiu and H. Tye (Gordon and Breach, 1984).

\bibitem{CD} C. Dunning: 
\emph{J. Phys.} \textbf{A36} (2003) 5463-5476, hep-th/0210225.

\bibitem{ferrara}S. Ferrara, L. Girardello and S. Sciuto: 
\emph{Phys.
Lett.} \textbf{B76} (1978) 303-306.

\bibitem{schoutens}K. Schoutens: 
\emph{Nucl. Phys.} \textbf{B344} (1990) 665-695.

\bibitem{heretic}T.R. Klassen and E. Melzer: \emph{Int. J. Mod. Phys.} \textbf{A8}
(1993) 4131-4174, hep-th/9206114.

\bibitem{FRTodd}G. Feverati, F. Ravanini and G. Takács: \emph{Phys. Lett.} \textbf{B444}
(1998) 442-450, hep-th/9807160.

\bibitem{polymer}A.B. Zamolodchikov: \emph{Phys. Lett.} \textbf{B335} (1994) 436-443.

\bibitem{folded}Z. Bajnok, L. Palla, G. Takács and F. Wágner:
  \emph{Nucl. Phys.} \textbf{B587} 
(2000) 585-618, hep-th/0004181.

\bibitem{HolM}T. Hollowood and A. Mavrikis: 
\emph{Nucl. Phys.} \textbf{B484} (1997) 631, hep-th/9606116.
%

\bibitem{Zamp}A.B. Zamolodchikov: \emph{Int. J. Mod. Phys.} \textbf{A3} (1988)
4235; \emph{Adv. Stud. Pure Math.} \textbf{19} (1989) 614. 

\bibitem{Mansf}H.C. Liao and P. Mansfield: \emph{Nucl. Phys.} \textbf{B344} (1990)
696. 

\bibitem{Ger}P. Mathieu: \emph{Nucl. Phys.} \textbf{B336} (1990)
338-348;\\
P. Mathieu and G.M.T. Watts: \emph{Nucl. Phys.}
\textbf{B510} (1998) 577-607, hep-th/9707050.

\bibitem{Inam}T. Inami, S. Odake and Y.Z. Zhang: \emph{Phys. Lett.} \textbf{B359}
(1995) 118-124, hep-th/9506157;\\
 R.I. Nepomechie: \emph{Phys.
Lett.} \textbf{B509} (2001) 183-188, hep-th/0103029. 

\bibitem{BF}P. Baseilhac and V.A. Fateev: \emph{Nucl. Phys.} \textbf{B532} (1998)
567-587, hep-th/9906010. 

\bibitem{KM}T.R. Klassen and E. Melzer: \emph{Nucl. Phys.}
  \textbf{B350} (1991) 635-689.

\bibitem{yang}
C.N.\ Yang and C.P.\ Yang:
\emph{J. Math. Phys.} \textbf{10}  (1969) 1115-1122.

\bibitem{ZamTBA}
Al.B.\ Zamolodchikov:
\emph{Nucl. Phys.} \textbf{B342} (1990) 695-720.

\bibitem{FI}
P.\ Fendley and K.\ Intriligator:
\emph{Nucl. Phys.} \textbf{B372} (1992) 533-558, hep-th/9111014.

\bibitem{RTV1}
F.\ Ravanini, R.\ Tateo and A.\ Valleriani:
\emph{Int. J. Mod. Phys.} \textbf{A8}  (1993) 1707-1727, hep-th/9207040. 

\bibitem{RTson}
R.\ Tateo:
\emph{Int. J. Mod. Phys.} \textbf {A10} (1995) 1357-1376, hep-th/9405197.

\bibitem{RT}
R.\ Tateo:
\emph{Phys. Lett.} \textbf{B355}  (1995) 157-164, hep-th/9505022.

\bibitem{DDV}
C.\ Destri and H.J.\ de Vega:
\emph{Phys. Rev. Lett.} \textbf{69} (1992) 2313-2317;\\
\emph{Nucl. Phys.} \textbf{B438}  (1995) 413-454, hep-th/9407117.

\bibitem{KP}
A.\ Kl\"umper and P.A.\ Pearce:
\emph{J. Stat. Phys.} \textbf {64} (1991) 13-76.

\bibitem{KBP}
A.\ Kl\"umper, M.T.\ Batchelor and P.A.\ Pearce:
\emph{J. Phys.} \textbf{A24} (1991) 3111-3133.

\bibitem{DTan1}
P.\ Dorey and R.\ Tateo: 
\emph{Nucl. Phys.} \textbf{B482 }  (1996) 639-659, hep-th/9607167.

\bibitem{DTan2}
P.\ Dorey and R.\ Tateo:
\emph{Nucl. Phys.} \textbf{B515}  (1998) 575-623, hep-th/9706140.

\bibitem{FRT3}
G.\ Feverati, F.\ Ravanini and  G.\ Takács:
\emph{Nucl. Phys.} \textbf{B570}  (2000) 615-643, hep-th/9909031. 

\bibitem{FMQR}
D.\ Fioravanti, A.\ Mariottini, E.\ Quattrini and F.\ Ravanini:
\emph{Phys. Rev. Lett.} \textbf{B390} (1997) 243-251, hep-th/9608091.

\bibitem{DDV2}
C.\ Destri and H.J.\ de Vega:
\emph{Nucl. Phys.} \textbf{B504}(1997) 621-664, hep-th/9701107.

\bibitem{FRT2}
G.\ Feverati, F.\ Ravanini and G.\ Takács:
\emph{Phys. Lett.} \textbf{B430} (1998) 264-273, hep-th/9803104.

\bibitem{FRT1}
G.\ Feverati, F.\ Ravanini and G.\ Takács:
\emph{Nucl. Phys.} \textbf{B540} (1999) 543-586, hep-th/9805117.

\bibitem{TCSA}
V.~P.~Yurov and A.~B.~Zamolodchikov:
\emph{Int. J. Mod. Phys.} \textbf{A5} (1990) 3221;\\
V.~P.~Yurov and A.~B.~Zamolodchikov:
\emph{Int. J. Mod. Phys.} \textbf{A6} (1991) 4557.

\bibitem{BSG}
Z.~Bajnok, L.~Palla and G.~Takács,
\emph{Nucl. Phys.} \textbf{B614} (2001) 405, hep-th/0106069; \\
Z.~Bajnok, L.~Palla and G.~Takács,
\emph{Nucl. Phys.} \textbf{B622} (2002) 565, hep-th/0108157. 

\bibitem{BSSG}
Z.~Bajnok, L.~Palla and G.~Takács,
\emph{Nucl. Phys.} \textbf{B644} (2002) 509, hep-th/0207099.

\end{thebibliography}
\end{document}